**TITLE:**

Electron FLASH platform for pre-clinical research: LINAC modification, simplification of pulse control and dosimetry

**AUTHORS:**

Banghao Zhou[1&*], Lixiang Guo[1&], Weiguo Lu[2], Mahbubur Rahman[2], Rongxiao Zhang[3], Varghese Anto Chirayath[4], Yang Kyun Park[2], Strahinja Stojadinovic[2], Marvin Garza[2], and Ken Kang-Hsin Wang[1*]

[1] Biomedical Imaging and Radiation Technology Laboratory (BIRTLab), Department of Radiation Oncology, University of Texas Southwestern Medical Center, Dallas, Texas 75235, USA

[2] Department of Radiation Oncology, University of Texas Southwestern Medical Center, Dallas, Texas 75235, USA

[3] Department of Radiation Medicine, New York Medical College, Valhalla, NY 10595

[4] Department of Physics, College of Science, The University of Texas at Arlington, Arlington, TX 76019, USA

[&] These authors contributed equally.

**\*CORRESPONDENCE:**

Banghao Zhou

Department of Radiation Oncology, University of Texas Southwestern Medical Center, 2280 Inwood Road, Dallas, TX 75235, USA

Tel: 469-660-6176

Email: banghao.zhou@utsouthwestern.edu

Ken Kang-Hsin Wang, PhD

Department of Radiation Oncology, University of Texas Southwestern Medical Center, 2280 Inwood Road, Dallas, TX 75235, USA

Tel: 614-282-0859

Email: kang-hsin.wang@utsouthwestern.edu

**RUNNING TITLE:**

Research platform for electron FLASH

**ABSTRACT**

**Background**: FLASH radiotherapy (FLASH-RT) is a treatment regime that delivers therapeutic dose to tumors at an ultra-high dose rate (UHDR, > 40 Gy/s) while maintaining adequate normal tissue sparing. However, a comprehensive understanding of the underlying mechanisms, potential late toxicities, and optimal fractionation schemes is important for successful clinical translation. This has necessitated extensive pre-clinical investigations, leading several research institutions to initiate dedicated FLASH research programs.

**Purpose**: This work describes a workflow for establishing an easily accessible electron FLASH (eFLASH) platform. The platform incorporates simplified pulse control, optimized dose rate delivery, and validated Monte Carlo (MC) dose engine for accurate in vivo dosimetry dedicated to FLASH pre-clinical studies.

**Methods**: A Varian linear accelerator (LINAC) converted from 6 MV X-ray to 6 MeV eFLASH by retracting the target and flattening filter, was used for the study. FLASH pulse delivery was controlled either by monitor unit (MU) setting or a simplified external controller. Adjustment of the automatic frequency control (AFC) module allowed us to optimize the LINAC pulse form to achieve a uniform dose rate throughout FLASH irradiation. EBT-XD films were used for beam data measurements. A MC model for the 6 MeV FLASH beam was commissioned to ensure accurate dose calculation necessary for reproducible in vivo studies. To illustrate the significance of accurate dose calculation in quantifying dose inhomogeneities for FLASH pre-clinical studies, we employed in vivo mouse whole brain and rat spinal cord irradiations as case studies.

**Results**: Optimizing the AFC module enabled the generation of a uniform pulse form, ensuring consistent dose per pulse (DPP) and a uniform dose rate throughout FLASH irradiation. The 6 MeV FLASH beam can sustain dose rates exceeding 40 Gy/s for depths close to 2 cm, with field sizes ranging from a 10×10 cm$^2$ to a 1 cm diameter circular field, sufficient for small animal studies. The MC model closely agreed with film measurements, with an average absolute difference within 2% for all percent depth dose (PDDs) and profiles at various field sizes. MC dose calculations indicated that 6 MeV FLASH is adequate to achieve a uniform dose distribution for mouse whole brain irradiation but may not be optimal for the spinal cord study.

**Conclusions**: Leveraging published studies, we present a novel workflow for establishing a LINAC-based eFLASH research platform. This approach incorporates techniques for optimized dose rate delivery, a simplified pulse control system, and validated MC engine specifically tailored for FLASH pre-clinical studies. This work provides researchers with valuable new approaches to facilitate the development of robust and accessible LINAC-based system for FLASH studies.

## 1. INTRODUCTION

FLASH radiotherapy (FLASH-RT) is a treatment regimen enabling the delivery of curative dose to tumors at an ultra-high dose rate (UHDR, > 40 Gy/s), while preserving adequate normal tissue protection — a phenomenon known as the FLASH effect[1]. Despite its considerable potential for clinical practice, researchers have not fully understood the underlying mechanisms of the FLASH effect, late toxicity, and fractionation scheme. This has necessitated extensive pre-clinical investigations, leading several research institutions to initiate dedicated FLASH research programs.

UHDR delivery can be administered through various modalities, including photon, proton or heavy charged particles, and electron. Photon UHDR was achieved using kilovoltage (kV) X-ray tubes[2-5], megavoltage (MV) photons from linear accelerators (LINACs)[6-9], and synchrotron light sources producing kV X-rays[10,11]. Proton or heavy charged particles have also been employed for UHDR delivery[12-19]. However, the existing pre-clinical photon UHDR systems utilizing X-ray tubes face limitations in dose rate and operating space, making them less effective for treating deep-seated targets and large animals[2-5]. The utilization of clinical LINACs for photon UHDR study is also compounded by the low efficiency in Bremsstrahlung generation and the limited beam currents. Moreover, synchrotron-based photon UHDR systems and proton UHDR irradiators are not widely accessible, restricting their use to a relatively small group of researchers.

In comparison, electron FLASH (eFLASH) offers a more accessible approach through the modification of widely used clinical LINACs. A Varian Clinac 21EX was adapted for UHDR irradiation by Schüler et al.[20] and was used for mice study[21]. Lempart et al.[22] performed modifications and fine-tuning on an Elekta Precise LINAC to enable eFLASH irradiation. Following this, Xie et al.[23] simplified the modification process and converted an Elektra Synergy LINAC into eFLASH mode for conducting in vivo experiments on mice. However, in these studies, UHDR beams were generated within the LINAC head or at distances of less than 53 cm from the target, limiting the application of UHDR for large animals and patient treatments. Rahmen et al.[24] further contributed by modifying a Varian Clinac 2100 C/D to facilitate UHDR deliver at the treatment room isocenter. Recently, No et al. also explored the feasibility of applying LINAC-based eFLASH for clinical trials[25]. Byrne et al. commissioned a high-throughput, variable dose rate eFLASH platform for murine whole-thoracic lung irradiation study based on a Varian 21EX LINAC[26]. These advancements facilitated the use of eFLASH for research and clinical translation.

To optimize the efficacy of pre-clinical studies using the eFLASH research platform, it is imperative to address several specific aspects. These improvements aimed to enhance the accessibility and operation of the eFLASH platform, ensure accurate and reproducible UHDR delivery, and provide accurate in vivo dosimetry to facilitate the interpretation of experimental outcomes. A ramp-up time within the initial 4 to 6 pulses was observed by Rahman[24], resulting in a delay in reaching a stable dose per pulse (DPP) and, consequently, inconsistent dose rate throughout the irradiation. Additionally, some pulse control systems in previous studies required an optocoupler circuit[22] or a dedicated gating switchbox[24] to terminate the beam after reaching the desired pulse number, increasing system complexity and limiting accessibility. Furthermore, while pre-clinical FLASH studies have expanded in recent years, most relied on phenomenological observations on animals, with limited reporting on in vivo dosimetry details. The concerns regarding in vivo dose inhomogeneities have either been unacknowledged or acknowledged but not addressed. To resolve this need, Rahman proposed a treatment planning system for eFLASH using Eclipse based on GAMOS Monte Carlo (MC) package[27] to investigate FLASH dosimetry.

In this work, we presented a detailed workflow for establishing a pre-clinical LINAC-based eFLASH research platform. We achieved the consistency of DPP and dose rate by implementing a pulse optimization method, adjusting the automatic frequency control (AFC) module. We proposed a modified pulse control system by utilizing the external customer-defined dosimetry interlock (CDOS) commonly established in Varian Clinac LINACs to terminate the beam without the need of additional apparatus. Building on the previous work of Rahman et al.[27], we developed and refined a MC calculation engine tailored for eFLASH animal studies, providing accurate in vivo dosimetry. We present our work based on 6 MeV, but our workflow can be readily extended to other energy modes.

The spatial dose distributions or FLASH beam data were characterized with radiochromic films for the MC engine. Planning studies of mouse whole brain and rat spinal cord irradiation computed by the MC engine are presented in this work. The whole brain case is commonly used in FLASH studies[10,21,28], and the spinal cord is one important site in assessing how FLASH would spare late responding organ for clinical translation.

These planning studies demonstrate the heterogeneity of electron dosimetry and underscore the necessity for accurate dose calculation for in vivo FLASH studies. Our established workflow would allow investigators to establish an easily accessible eFLASH platform with a simplified pulse control system, optimized dose rate delivery, and an accurate MC engine for pre-clinical studies. We expect this platform will provide researchers with the opportunities to investigate underlying biological mechanisms of FLASH-RT and accelerate its translation into clinical practice.

## 2. METHODS AND MATERIALS

### 2. A. Machine modification

Adopting the methodology outlined by Rahman et al.[24], we converted a Varian 21EX LINAC into an eFLASH irradiator by configuring the LINAC in photon mode and retracting the target and the carousel in an empty port position. In conventional photon mode, the high-intensity electron beams hitting the target are converted into X-ray beams through Bremsstrahlung, with an efficiency of approximately 20-40%[29] for 6-18 MeV electrons. The flattening filters on the carousel can reduce the dose rate by 2-4 times[30]. By removing both the target and flattening filter from the beam path, we can utilize the maximum electron beam fluence to achieve UHDR irradiation. In the conventional operation of the LINAC, essential components such as the target, carousel, and energy switch are pneumatically controlled through air valves. To deliver the high-intensity UHDR electron beams via photon mode, the air drive in the LINAC backstand was disabled in the first step, allowing us to manually adjust and secure the positions of these components. An empty port on the carousel was used and the X-ray target was mechanically removed, ensuring no flattening filter or target was within the electron beam path. Depending on the energy of choice, we mechanically locked the energy switch at the corresponding position to prevent shifting during FLASH irradiation. This setup enabled us to achieve the UHDR exceeding 40 Gy/s at the isocenter with the 6 MeV FLASH beam.

### 2. B. Pulse monitoring and control

#### 2. B. 1. Pulse form monitoring

To ensure accurate and reproducible UHDR dose delivery, it is essential to precisely control the number of pulses delivered and monitor the amplitudes of these pulses in real time. For real-time monitoring purpose, we employed a pulse form monitoring unit (unit 1 in Fig. 1), which consists of an HC-120 series photomultiplier tube (PMT, Hamamatsu, Shizuoka Prefecture, Japan) coupled with an optical fiber. This unit, positioned outside the radiation field, detected scattered radiation through radiation-induced Cherenkov and fluorescence emission. The PMT output was fed into an oscilloscope (70MHz, PicoScope 3000 series, Pico Technology, Cambridgeshire, UK), allowing us to monitor the LINAC pulse amplitudes on a pulse-by-pulse basis.

### 2. B. 2. MU and external pulse control

The precise delivery of pulses in FLASH irradiation can be controlled either by configuring the monitor unit (MU) at the LINAC console or by utilizing an external pulse control system. In conventional mode, the LINAC employs monitor chambers to monitor various beam parameters, such as dose, mean dose rate, DPP, beam steering, and positioning[31]. However, under high DPP conditions at UHDR, the chambers may exhibit saturation or decreased ion collection efficiency, presenting challenges for online dose monitoring[32]. Despite this, it remains feasible to control the pulse number by setting MU at the LINAC console with limitation of adjusting desired number of pulse and therefore restricting the resolution of dose achieved by the eFLASH, as shown in the later results.

While MU settings offer limited options for pulse delivery, an additional external pulse control unit (unit 2 in Fig. 1) was developed to offer greater flexibility as a complementary method. The number of pulses was counted by a remote trigger unit (RTU, DoseOptics LLC, NH) positioned outside the radiation field. This RTU, a coincidence-based radiation detector, features two scintillators and corresponding silicon photomultipliers (SiPM) for detecting scattered radiation. An Arduino-RTU (Arduino, Turin, Italy) circuit was designed to count the pulses and to deactivate a gating reed relay once the desired pulse count was reached. This reed relay was connected across pins 1 and 2 of J15 on the stand mother printed circuit board (PCB) at the back of our LINAC. Disconnection of these pins triggers the CDOS interlock established in the Varian LINAC system (unit 3 in Fig. 1), leading to the immediate cessation of the radiation beam.

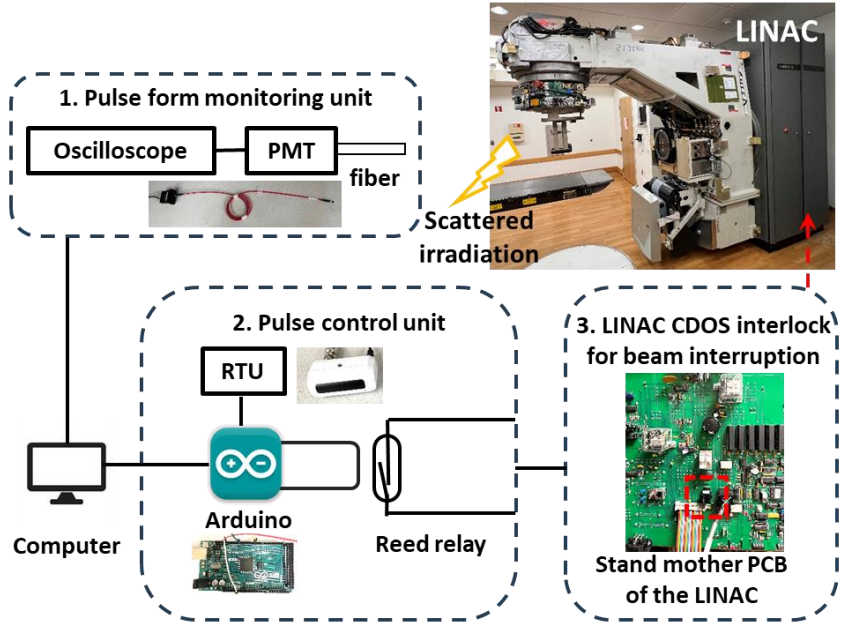

**Figure 1**. A schematic view shows the eFLASH and the (1) external pulse form monitoring unit and (2) external pulse control unit for monitoring and controlling FLASH pulses through scattered irradiation. The pulse control is achieved through the (3) CDOS jumper (red dash in the stand mother printed circuit board PCB) to trigger LINAC interlock and therefore beam interruption.

## 2. C. Pulse form optimization

In our standing wave accelerator, the microwave power must be supplied at the resonant frequency of the accelerating guide. This is managed by the AFC system in the LINAC, which adjusts the frequency of the radiofrequency (RF) driver based on the phase relationship between RF forward and reflected power samples from the waveguide[31]. However, the LINAC's automatic feedback mechanisms were unable to fine-tune the beam within the short span of the initial several pulses in our 6MeV FLASH mode, resulting in several unstable pulses with varying amplitudes. These variations can lead to inconsistent DPP and dose rate, and further introduce dosimetry uncertainties in UHDR delivery. To address this issue, we devised an optimization strategy to adjust the pulse form for our 6 MeV FLASH beam by disabling the AFC servo and manually aligning the microwave source frequency from the RF driver with the waveguide's resonant frequency using the AFC manual mode. We used the PMT signal to monitor the pulse until a stable pulse form was achieved during the adjustment.

## 2. D. Pulse form analysis

Each pulse in the waveform exhibits a characteristic ramp-up and decay time, with an approximate duration of 4.5 µs. The area under the curve (AUC) of these pulses was found to correlate with the delivered dose. We observed that the total AUC of all pulses, after subtracting the background signal, showed a covariance close to 1 with the total dose measured by the film. This relationship enables us to calculate the dose delivered by each pulse using the ratio of its AUC to the total AUC, multiplied by the total delivered dose. With this method, we can analyze the DPP distribution in both AFC on and off modes based on the waveform and estimate how much dose is delivered at various dose rates.

## 2. E. Beam data measurement

We employed Gafchromic EBT-XD films (Ashland, Bridgewater, NJ) for our FLASH beam data measurement and MC engine commissioning, as these films have demonstrated superior dose rate and energy independence[33-35]. The films were scanned 24 hours post-irradiation using a photo flatbed scanner (Epson Expression 12000XL, Suwa, Nagano, Japan). An in-house film analysis software was employed to extract the dose information from films, utilizing a two-channel method (green and blue)[36]. The film dosimetry was validated using optically stimulated luminescence dosimeters (OSLDs), with differences within 4%. Our measurements encompassed percent depth dose (PDD), dose profiles, dose rates, and output factors.

For beam data measurements, films were positioned orthogonal to and centered along the beam's central axis (CAX), either at the surface or at various depths between 30×30 cm² solid water slabs, with a backscatter piece at 6 cm thickness. Our measurements were conducted at 100 cm source-to-surface (SSD) by default. The beam data measurements were conducted for various field sizes, including open field, 10×10 cm², 6×6 cm², and the circular fields at diameters of 3, 2, 1.5, and 1 cm within the 6×6 cm² cone.

For PDD measurements, we obtained the depth dose along the CAX by averaging the dose within the central 0.3×0.3 cm² area for 1 and 1.5 cm diameter circular fields, or 0.5×0.5 cm² area for other fields, of 2.5×2.5 cm² square films placed at various depths at SSD 100 cm. The film-measured PDDs were fitted with a Fermi-Dirac distribution multiplied by a third-degree polynomial with coefficient of determination ($R^2$) ≥ 0.99 to generate continuous PDD curves.

Crossline and inline profiles were measured using 25×2.5 cm² films in solid water at 100 cm SSD at surface, and depths of 0.8, 1.3, 1.8, and 2.3 cm. The profiles at 1.3 cm depth were also measured at other SSDs of 95, 110, and 120 cm. Furthermore, the in-air profiles at 95 cm SSD were also measured by placing the films on the surface of Styrofoam blocks.

The output factor for a given field size was determined based on the film measurement, relative to the dose of 10×10 cm² field size at depth of maximum dose ($d_{max}$). The dose rates, DPP, and instantaneous dose rates were calculated based on the delivered dose, pulse number, pulse repetition frequency (PRF), and pulse width information through our pulse monitoring and control units (Fig. 1).

## 2. F. FLASH beam modeling and dose calculation

### 2. F. 1. MC beam modeling

It is widely recognized that accurate dose calculations for electron beams often require the use of MC simulations[37]. Considering this, we have developed a Geant4-based GAMOS[38] MC dose calculation engine tailored for pre-clinical studies. It is crucial to accurately model all the necessary components in the LINAC head that are traversed by an electron beam to determine its detailed characteristics[39]. This is because electrons can easily scatter and generate X-rays by interacting with the LINAC head components. We explicitly modeled the LINAC head geometry using the GAMOS package, including the target, primary collimator, flattening filter or scattering foil, Beryllium window, monitor chambers, field light mirror, shielding materials, jaws, electron applicators and cutouts. For the eFLASH mode, the target and flattening filter were removed from the LINAC geometry in X-ray mode (Fig. 2), corresponding to our machine modification procedure. Based upon the work of Rahman et al.[27], we introduced enhancements in our model. We modified the representation of the X jaws, modeling them as moving in a linear trajectory instead of the simplified arc trajectory. The Y jaws were kept the same as they move along an arc trajectory in reality. Furthermore, we refined the modeling of the electron applicators, representing them with three layers, as opposed to a simplistic single bottom layer.

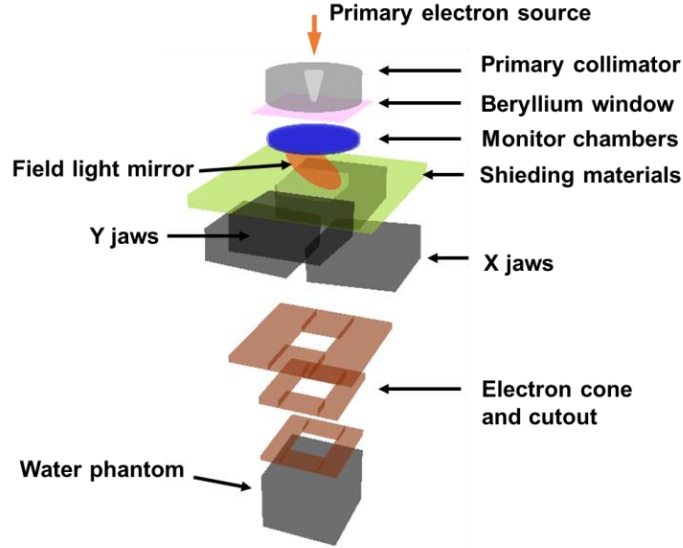

**Figure 2**. Setup for MC simulations of the eFLASH beam.

Given the LINAC head geometry (Fig. 2), to extract the necessary beam parameters for the FLASH beam, we need to determine the source emission parameters including mean energy ($E$), mean energy spread ($\sigma_E$), source emittance cone angle ($\theta_{cone}$), and spot size ($\sigma$) through a beam commissioning process.

The MC model commissioning was performed using the open-field setting, where the jaws were opened at around 40×40 cm² field size and electron applicators were removed. This setting is sensitive to beam and geometry parameters, making it effective to extract the beam source parameters[40, 41]. To obtain PDD and profiles, we adopted a simulated 30×30×8 cm³ water phantom, composing of 0.1×0.1×0.1 cm³ voxels and a particle number of 2×10⁸ in our MC modeling.

In the initial step of beam parameter tuning, where PDD fitting was involved, we firstly varied $E$ from 5.50 to 6.50 MeV and $\sigma_E$ from 0.1 to 0.9 MeV. The σ was kept constant at 0.5 mm, a value shown to effectively represent the electron beam[42], and $\theta_{cone}$ was initially assumed to be 0°. The optimal combination of $E$ and $\sigma_E$ was determined by achieving the lowest average absolute difference (AAD) < 2% between the MC-simulated PDD and the film-measured PDD under the open-field setting. Based on the optimal values of $E$ and $\sigma_E$, we then varied $\theta_{cone}$ from 0.0 to 8.0° while keeping σ constant at 0.5 mm for profile fitting. Given the open-field setting, the optimized $\theta_{cone}$ was selected when the AAD < 2% between MC-simulated and film-measured profiles can be achieved along both crossline and inline directions at surface, and depths of 0.8, 1.3, 1.8 and 2.3 cm. Moreover, the optimal $\theta_{cone}$ was also confirmed by comparing the simulated in-air profiles at 95 cm SSD and profiles at 1.3 cm depth in solid water at SSDs of 95, 110, and 120 cm to those of measurements, with AAD < 2% for each comparison.

Finally, the beam model with the optimized parameters established from the open-field setting was validated by the measured PDD and profiles at field sizes commonly used in pre-clinical setting including 10×10 cm², 6×6 cm², and circular fields of 3, 2, 1.5, and 1 cm diameters. The setting for MC simulation used in the validation step was the same as that of beam modeling process except various electron applicators, cutouts, and corresponding jaw opening were applied. The validation criterion was that the AAD between MC simulations and film measurements should be < 2% for all the PDD and profiles.

### 2. F. 2. MC dose calculation in vivo

After the 6 MeV FLASH beam model was established, we developed a MC dose engine tailored for pre-clinical studies. Phase space (PS) files at the plane of 96 cm SSD were generated for the abovementioned field sizes and were subsequently utilized for dose calculations in objects placed below electron cones. The absolute output of the 6 MeV FLASH beam with a 10×10 cm² field at $d_{max}$ was determined by averaging the dose from three film measurements and was used as the calibration point for absolute dose calculation in the MC engine.

For in vivo dose calculations, cone beam computed tomography (CBCT) scans of the animals were conducted using either X-RAD 225 (Precision X-Ray, Madison, CT) or small animal radiation research platform (SARRP, Xstrahl, Surrey, UK) irradiators. The acquired CBCT scans were imported into Eclipse (Varian, Palo Alto, CA) for tissue segmentation, and subsequently exported to GAMOS MC package. Material types based on the segmentation were assigned to the corresponding voxels of the CBCT images. Accordingly, a GAMOS-specific CT geometry file in "g4dcm" format was generated for subsequent in vivo dose calculation. It contains information regarding material types and distributions, geometry, and positioning of studied objects, used for computing radiation interactions with matter for dose calculation. The calculated dose was saved in binary files by default to expedite data processing. These files were later converted into the DICOM RTDOSE format, which can be visualized and analyzed in 3D Slicer[43] in conjunction with MATLAB (MathWorks, Natick, MA).

Following the process described above, we performed planning studies of in vivo cases with the 6 MeV FLASH MC model to illustrate the importance of having an accurate dose engine for pre-clinical research. The uncertainty of the MC dose calculation was maintained less than 3%. All animal procedures were carried out in accordance with the institutional animal care and use committee at the University of Texas Southwestern Medical Center. The first planning case involved a mouse whole brain irradiation (WBI) with a 1.7 cm diameter circular graphite aperture at 100 cm SSD, the similar experimental setup described in the work of Montay-Gruel et al.[28]. A single posterior-anterior (PA) beam was utilized. A CBCT scan of a C57BL/6J albino mouse (The Jackson Laboratory, Bar Harbor, ME) acquired by SARRP was used for the calculation. The dose distribution was normalized to the point dose at a depth of 0.25 cm along the beam CAX, around the center of the brain (the white dot in Fig. 6a2). For the second case, we considered rat C2-T2 spinal cord irradiation. A CBCT scan of a CD IGS rat (Charles River Laboratories, Wilmington, MA) was acquired using the X-RAD 225 and subsequently imported into GAMOS for dose calculation. A single PA beam with a 2×1 cm$^2$ field size at 100 cm SSD was designed for the spinal cord irradiation. The dose distribution was normalized to the central point of the C2-T2 spinal cord (the white dot in Fig. 6b2).

## 3. RESULTS

### 3. A. Pulse control and optimization

The pulse form of the 6 MeV FLASH beam with the LINAC default AFC setting is shown in Fig. 3a, where an initial ramp-up time was observed. Due to this ramp-up phase, several pulses exhibited varying amplitudes, indicating inconsistent DPP and dose rate, which could potentially lead to dosimetric uncertainties during FLASH irradiation. In contrast, the waveform after the AFC optimization is shown in Fig. 3b, where the pulse amplitudes remained relatively stable, representing stable DPP and dose rate. Figure 3c illustrates mean dose rates versus the number of pulses delivered with and without AFC optimization, measured at solid water surface and field size of 10×10 cm$^2$ at 100 cm SSD. Without optimization, the dose rate varied from 20 to 220 Gy/s within the first 60 pulses, corresponding to approximately 40 Gy delivery, which covers most dose levels used in pre-clinical studies. In contrast, after optimization, a constant UHDR around 220 Gy/s could be maintained. If one would deliver a fixed dose, i.e. 10.7 Gy around the dose level at which the FLASH effect starts to emerge[44], the corresponding DPP vs. number of pulses with and without AFC optimization is shown in Fig. 3d. The corresponding dose delivered within various dose rate ranges with and without AFC optimized are shown in Fig. 3e and f, respectively. With the default AFC setting, 20 pulses were needed to deliver the 10.7 Gy. We observed that the DPP increased from 0.005 to 0.93 Gy from the $1^{st}$ pulse to the $14^{th}$ pulse and then reached a stabilizing phase (Fig. 3d). This resulted in 20% of the total dose being delivered with a non-uniform dose rate, and even 2% of the dose was delivered $\leq 40$ Gy/s, falling outside the FLASH dose rate range (Fig. 3e). After the AFC optimization, only 17 pulses were needed to deliver the 10.7 Gy and the DPP remained stable around 0.6 Gy/pulse throughout the entire irradiation. This ensured that the dose was delivered within a tight dose rate range 200 to 240 Gy/s (Fig. 3f). These results demonstrate that the AFC optimization can support consistent FLASH delivery. Figure 3g demonstrates that the number of delivered pulses correlates linearly with the MU setting, with each MU corresponding to 6 or 7 pulses, measured at the solid water surface and $10\times10$ cm$^2$ field. Although it offers limited options in terms of pulse numbers, MU setting allows us to deliver selected dose. As a complementary method, the external control unit (unit 2, Fig. 1) would miss 1-3 pulses beyond the set pulse number (Fig. 3h). This discrepancy might arise from the limited response speed of the Arduino circuit or the reed relay, which could be improved by either optimizing the circuit or utilizing solid-state relays.

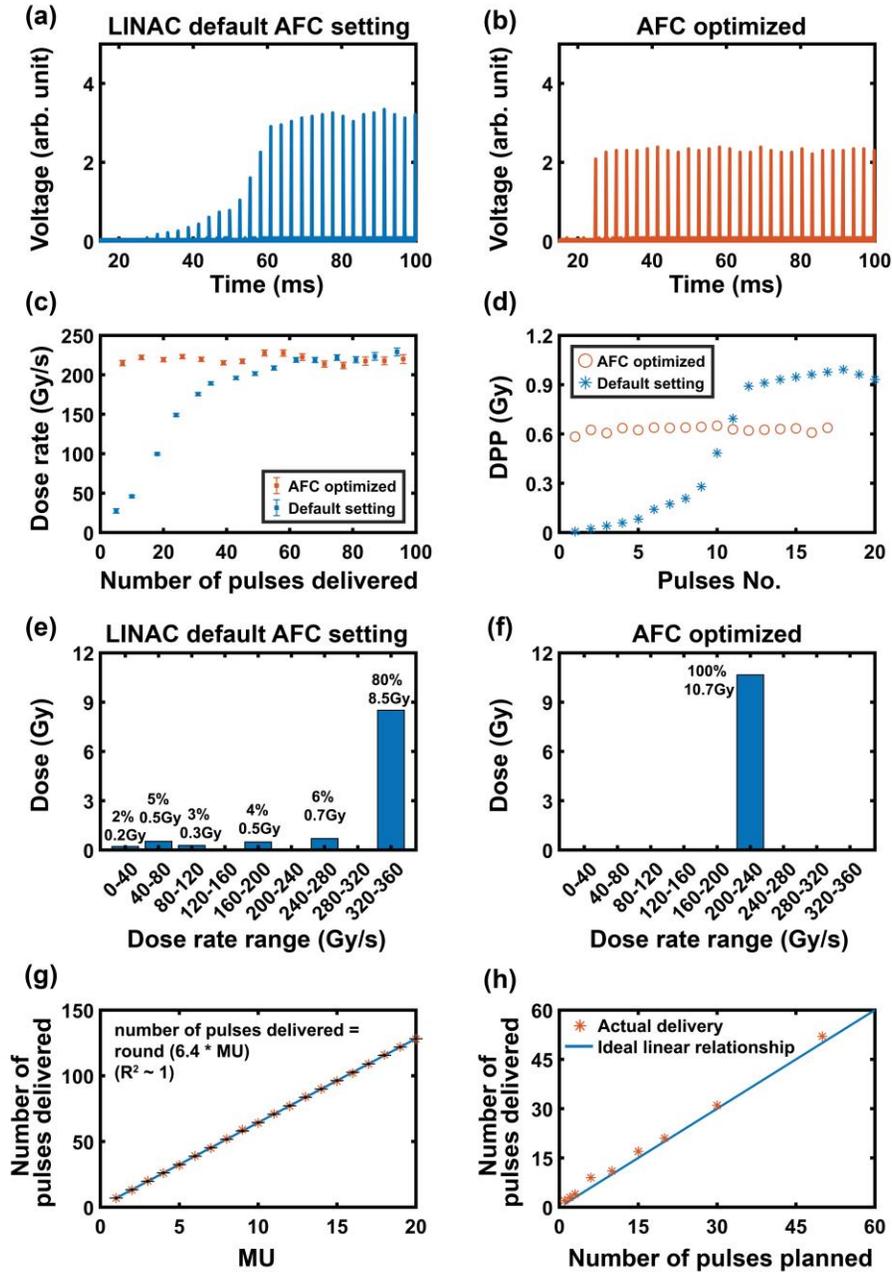

**Figure 3.** (a) and (b) show the representatives of pulse waveform versus the pulse delivered time for the default AFC setting and after AFC optimization, respectively; (c) shows the mean dose rate versus the number of pulses delivered and (d) is the corresponding DPP versus individual pulse for 10.7 Gy based on the default AFC setting and AFC optimization for a field size of 10×10 cm$^2$ at 100 cm SSD; (e) and (f) illustrate the 10.7 Gy delivered at various dose rates ranges based on the default AFC setting and the AFC optimization, respectively; (g) shows the number of pulse delivered correlated linearly with the MU after the AFC optimization; (h) shows the number of pulse delivered versus planned in the external pulse control system.

### 3. B. Beam data

The film-measured and the Fermi-Dirac fitted PDD curves for the 6 MeV FLASH beam are shown in Fig. 4a. For the sake of comparison, the PDD of our conventional 6 MeV electron beam is also displayed. The FLASH beam, compared to the conventional beam, exhibits lower energy while having a similar $d_{max}$ around 1.3 cm. Figures 4b and c illustrate the FLASH PDDs and corresponding depth dose rates respectively at 100 cm SSD for a given field size. Given the pulse width of 4.5 μs and a PRF of 360 Hz for the 6 MeV FLASH, the surface dose rate under 10×10 cm² can reach 218 Gy/s at 100 cm SSD, corresponding to 0.6 Gy/pulse and an instantaneous dose rate at $1.34 \times 10^5$ Gy/s. The detailed beam characteristics at isocenter for each field size are shown in the supplementary Table S1. Along with the decrease of field size, the $d_{max}$ was shifted from 1.34 to 0.40 cm (Fig. 4b), while the dose rate at $d_{max}$ decreasing from 320 to 171 Gy/s (Fig. 4c). The $d_{max}$, corresponding dose rates, and output factors are shown in the supplementary Table S2 for all the field sizes. Figures 4d and e display the representative crossline relative dose and dose rate profiles, respectively, at field size of 10×10 cm². The 90% dose can be maintained within the central 8×8 cm², and the dose rate can be sustained at > 40 Gy/s within the entire 10×10 cm² even at 2.3 cm depth. The corresponding dose rate profiles of the 1 cm diameter circular field at crossline direction are shown in Fig. 4f, where the dose rate can be maintained > 40 Gy/s across the 1 cm field at a depth up to 1.8 cm. In sum, according to the results of Figs. 4c, e, and f, our 6 MeV FLASH beam could be used for small animal irradiation or superficial irradiation in large animals at depth close to 2 cm.

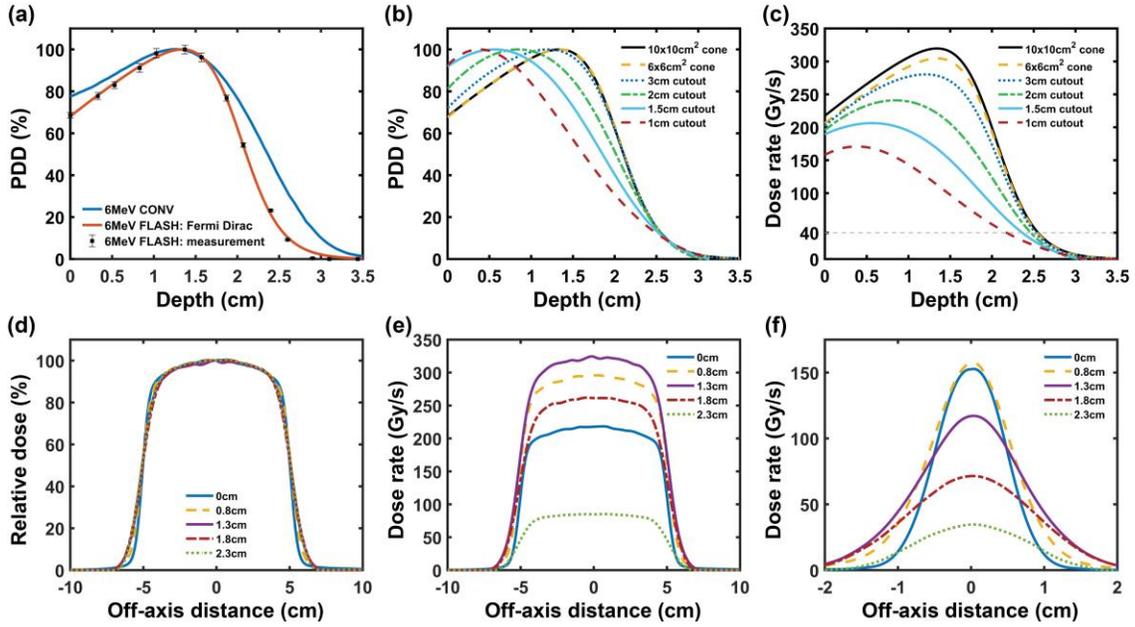

**Figure 4.** Measured 6 MeV FLASH beam characteristics: (a) FLASH vs. conventional 6 MeV PDD at field size of 10×10 cm²; (b) PDDs and (c) depth dose rates along CAX for field size of 10×10 cm², 6×6 cm² and 3, 2, 1.5, and 1 cm diameter circular fields based on a 6×6 cm² cone; (d) and (e) are the crossline profiles and dose rate profiles respectively, for 10×10 cm² field at depths of surface, 0.8, 1.3, 1.8 and 2.3 cm; (f) is the corresponding crossline dose rate profiles of 1 cm diameter circular field.

### 3. C. MC beam modeling

By the MC PDD modeling, we determined the optimal parameters $E = 5.53$ MeV and $\sigma_E = 0.3$ MeV, resulting in AAD of 1.23% between the MC-simulated and film-measured PDDs in the open-field scenario (Fig. 5a). These parameters were further validated for the other field sizes, each exhibiting AAD within 2%. For instance, Fig. 5b demonstrates the validation case for a 1 cm circular field within 1.31% accuracy, as this field being commonly used for pre-clinical studies. The PDD comparisons for other field sizes are shown in Fig. S1 in the supplementary material.

Given $E = 5.53$ MeV and $\sigma_E = 0.3$ MeV, taking σ as 0.5 mm, we determined the optimal $\theta_{cone}$ as 6.0° in the open-field setting. The optimal $\theta_{cone}$ led to AAD < 2% in the open-field setting between the computed profiles and the corresponding film measurements at depths from 0 to 2.3 cm along both crossline and inline directions (Fig. S2). Figure 5c shows a representative comparison at 1.3 cm depth with 0.84% AAD. Furthermore, the computed profiles at the same depth 1.3 cm at SSDs of 95, 110, and 120 cm and in-air profiles at 95 cm SSD also match to the corresponding film measurements, with AAD < 2% (Fig. S3). We further validated the beam model by comparing the computed and measured profiles for various depths and field sizes. Figure 5d displays the computed crossline profile at the 1.3 cm depth for the 1 cm circular field, with 1.49% AAD compared to the measurement. The comparisons for depths 0-2.3 cm in both crossline and inline directions for field sizes of 10×10 cm², and 6×6 cm², and circular fields at diameters of 3, 2, 1.5, and 1 cm are shown in Fig. S4 to S9 in the supplementary material, respectively. The AAD for each case is within 2%.

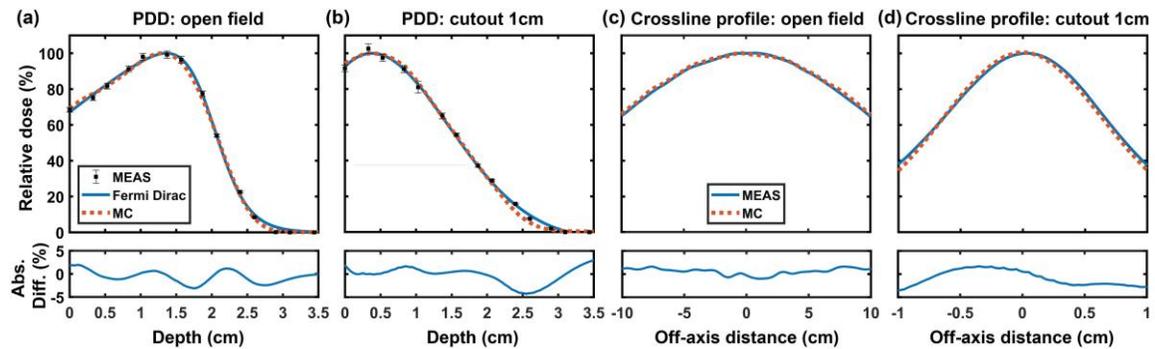

**Figure 5.** (a) MC-simulated PDD compared to film-measured PDD fitted by the Fermi-Dirac method for open field; (b) shows the case for the circular field at 1 cm diameter; (c) MC-simulated profile in comparison to film measurement at 1.3 cm depth open field; (d) shows the case of the 1 cm circular field. For each case, the difference between MC and measured PDDs or profiles is also shown.

### 3. D. Dose assessment for pre-clinical studies

The dose distribution for the mouse WBI study is shown in Figs. 6a1 and a2 for coronal and sagittal planes, respectively. Within 0.5 cm depth, approximately covering the thickness of the mouse brain, the PDD ranged from 90 to 100% (Fig. 6a3). The dose profiles showed > 90% along the head-to-tail direction and > 87% along the left-to-right direction within an off-axis distance (OAD) of -0.5 to 0.5 cm at the 0.25 cm depth, which is at middle position of the brain (Fig. 6a4). These results suggest the 6MeV FLASH beam with the 1.7 cm diameter circular cutout can achieve uniform dose distribution for mouse WBI.

Figures 6b1 and b2 illustrate the dose distributions for the case of rat C2-T2 spinal cord irradiation. While the 6 MeV FLASH beam can achieve more than 90% PDD at the depth of the C2-T2, it decreased rapidly beyond 2 cm (Fig. 6b3). Moreover, the profile in head-to-tail direction exhibited around 10% fall-off from the center to +/- 0.6 cm OAD (Fig. 6b4), resulting in a spinal cord length of < 0.8 cm not receiving a uniform dose. This might result in longitudinal dose-volume effect, which could escalate the median effective dose ($ED_{50}$, causing paresis in 50% of rats)[45], and thus increase the spinal cord dose tolerance. This dose-volume effect could confound the FLASH effect in sparing the spinal cord toxicity.

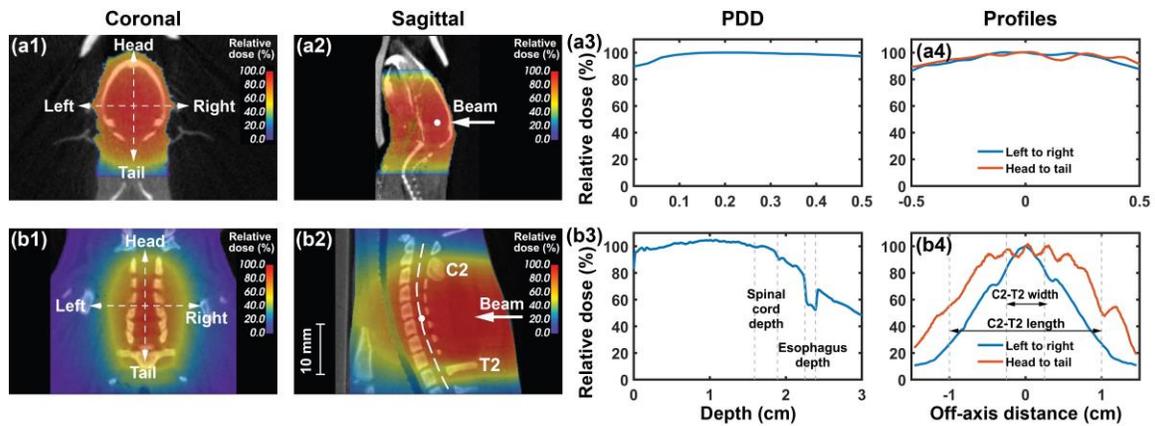

**Figure 6**. (a1-2) are the MC calculated dose distribution for 6 MeV FLASH in coronal and sagittal view, respectively, for mouse WBI with a 1.7 cm diameter graphite cutout. (a3-4) are the corresponding PDD along beam CAX and profiles, respectively. (b1-4) are the case for rat spinal cord irradiation with a 2×1 $cm^2$ field. The white dots in a2 and b2 are the dose normalization points.

## 4. DISCUSSION

Despite FLASH-RT has recently sparked tremendous interest in the radiation oncology community, the fundamental mechanisms driving the FLASH effect remain elusive. This knowledge gap has spurred the need for comprehensive pre-clinical research. Electron FLASH has been widely used for UHDR research[1,21,23,28,46-58], and the LINAC-based eFLASH has also been suggested for clinical trial use[25]. The significance of this work lies in the establishment of a workflow aiming to provide an easily accessible eFLASH platform with simple pulse control and accurate dosimetry system for in vivo studies.

Our innovative approach addressed the challenge of maintaining dosimetry consistency for UHDR irradiation by fine-tuning the AFC module, which ensures reproducible DPP and dose rates — critical parameters in FLASH research. Under the default AFC settings, a ramp-up period in the pulse waveform (Fig. 3a) was observed using the pulse monitoring system (Fig. 1). This initial period persisted for approximately the first 13 pulses within a beam delivery, resulting in inconsistent DPP and subsequently reducing the mean dose rate for a given pulse number. This effect was more pronounced when fewer pulses were delivered, as evidenced by mean dose rates of 27.4 Gy/s for the first 5 pulses and 45.8 Gy/s for the first 10 pulses. This suggests that a portion of dose was delivered at rates lower than 40 Gy/s, a threshold typically associated with the observation of the FLASH effect[1,44]. While it remains unclear whether the non-uniform DPP during delivery and the small amount of dose delivered at conventional rates impact the FLASH effect, ensuring a consistent dose rate and DPP throughout beam delivery is important for experiment reproducibility.

We further demonstrated the feasibility of utilizing MU setting to deliver the FLASH beam, as the given pulse or dose can be precisely quantified (Fig. 3g). Although the MU setting is limited by a fixed number of pulses, it provides a basic setup for initiating FLASH research without the need for a sophisticated pulse control system. Furthermore, we also introduced the use of CDOS interlock (Fig. 1) for pulse-by-pulse control for FLASH irradiation. As previously described, the advantage of utilizing the existing CDOS interlock is that it eliminates additional equipment for pulse control, i.e. optocoupler circuit[22] or gating switch box[24], which may not be readily available in some institutes. The initial results are encouraging, demonstrating that we can achieve control within 3 pulses (Fig. 3h). We believe that the accuracy of pulse control when using the CDOS setting could be improved by enhancing the response speed of the pulse control unit (unit 2 in Fig. 1), This enhancement could be achieved by either optimizing the speed of the circuit in gating pulse number detected by the RTU or by implementing solid-state relays with fast response.

While pre-clinical FLASH studies have expanded in recent years, most of them have relied on phenomenological observations in rodents, with limited reporting on in vivo dosimetry details. Moreover, in vivo dose inhomogeneities are not commonly reported. The development of an accurate dose calculation engine that generates in vivo plans for FLASH is pivotal for comprehending the underlying mechanism of UHDR. It is widely recognized that accurate dose calculations for electron beams often require the use of MC simulations[36]. Considering this, we have developed a GAMOS MC dose calculation system for our LINAC-based eFLASH. It is important to simulate accurately all the components in the LINAC head traversed by an electron beam to determine its detailed characteristics, because electron beams easily scatter and generate X-rays by interacting with the LINAC components. We explicitly modeled the head geometry under the GAMOS framework. For the eFLASH mode, the photon target was removed from the LINAC geometry (Fig. 2). The close agreement, < 2% of AAD, between the computed and measured PDDs and profiles suggests that our MC model accurately represents the 6MeV FLASH beam, which can be used for the in vivo dose calculations.

Multiple pre-clinical studies have been conducted to investigate the FLASH effect. The murine brain model has been commonly utilized[10,21,28,59-61]. Montay-Gruel et al. conducted mouse WBI research with a 6 MeV FLASH beam, revealing preserved spatial memory with mean dose rates exceeding 100 Gy/s[28]. Their subsequent investigations demonstrated that the 6 MeV FLASH can maintain microvasculature integrity[59] and limit reactive gliosis[60] in the brain compared to conventional radiotherapy. Alaghband et al. also conducted mouse WBI experiments with 6 MeV FLASH and observed neuroprotection provided by FLASH-RT in radiosensitive juvenile mice[61]. Beyond these phenomenological observations of FLASH effects, we employed MC dose calculation to inform the in vivo dosimetry for mouse WBI. Our results demonstrated that the 6MeV FLASH beam, using the similar setup as described in the study of Montay-Gruel et al.[28], was capable of delivering a uniform dose distribution to the mouse brain. This leads to the confidence in dose homogeneity when conducting WBI with low-energy eFLASH beams.

While most pre-clinical evidence supporting FLASH effects relies on acute toxicity, data for late-responding tissues are lacking. It is critical to determine whether FLASH can spare late-responding tissues before FLASH can be confidently translated to clinical practice. Given its steep dose-response relationship, the spinal cord serves as an ideal model to investigate if FLASH would increase the spinal cord dose tolerance and alter the response curve[45]. Rats have been used extensively to study radiation-induced injury to the spinal cord, exhibiting syndromes, paresis, similar to those observed in humans. Therefore, the rat spinal cord irradiation is of great interest to our group and scientific community to understand how FLASH would spare late responding organs. Bijl et al. demonstrated that when the irradiated length along the rat spinal cord was < 0.8 cm, the $ED_{50}$ for paresis increased dramatically with decreasing irradiated length[45], referred to as the dose-volume effect. The uniform dose distribution is an important consideration for the spinal cord study, as non-homogenous irradiation can potentially result in a shorter length of the spinal cord receiving the prescribed dose. This emphasizes the importance of having an accurate dose calculation engine for investigating in vivo dosimetry inhomogeneities. If one would use the 6 MeV FLASH beam for spinal cord irradiation, due to the dose fall-off approximately 10% along the head-to-tail direction from center to +/- 0.6 cm off-axis distance, a spinal cord length of < 0.8 cm would not receive a uniform dose. The longitudinal dose-volume effect could escalate the $ED_{50}$ and confound the outcome when assessing the FLASH sparing effect. Thus, higher-energy electrons are preferred for this case. These planning studies underscored the importance of utilizing accurate dose engine for electron FLASH studies, which can inform in vivo dose inhomogeneities and support study designs and interpretation of experiment outcomes.

## 5. CONCLUSION

We presented a streamlined eFLASH platform for pre-clinical studies, utilizing a modified LINAC system with simplified pulse control and optimized dose rate delivery. We also established a detailed MC dose engine closely matching the eFLASH beam. The consistent ultra-high dose rate throughout FLASH irradiation along with the detailed MC dose engine are expected to support study reproducibility. This work provides researchers with valuable new approaches to facilitate the development of robust and accessible LINAC-based system for FLASH research.

## ACKNOWLEDGEMENT

We acknowledge the funding support from Cancer Prevention and Research Institute of Texas, RR200042.

## CONFLICTS OF INTEREST STATEMENT

The authors have no relevant conflicts of interest to disclose.